\documentclass[12pt]{amsart}
\usepackage{geometry} 
\usepackage{amssymb}
\usepackage{amsmath}
\usepackage{graphicx}
\usepackage{multirow}
\usepackage{mathabx}
\usepackage{bigstrut}
\usepackage{lineno}
\usepackage{epstopdf}

\begin{document}

\title{On the Analytic Estimation of Radioactive Contamination 
from Degraded Alphas}
\author{Richard W. Kadel \\
{\fontsize{.25cm}{.07cm}\selectfont \emph{P\MakeLowercase{hysics }D\MakeLowercase{ivision}\\
L\MakeLowercase{awrence} B\MakeLowercase{erkeley} N\MakeLowercase{ational} L\MakeLowercase{aboratory (retired)}}}}
\date{6 December 2015\\  \phantom{Xx}email:  rwkadel@lbl.gov} 
\maketitle

\begin{abstract}
   The high energy spectrum of alpha particles emitted from a single isotope uniformly contaminating a bulk solid has a flat energy spectrum with a high end cutoff energy equal to
the maximal alpha kinetic energy ($T_{\alpha}$) of the decay. In this flat region of the spectrum, we show the surface rate $r_b\text{\,(Bq/keV-cm}^{2})$ arising from a bulk alpha contamination $\rho_b$ (Bq/cm$^3$) from a single isotope is given by  $r_b =\rho_b \Delta R/ 4 \Delta E $, where  $\Delta E = E_1-E_2>0\ $ is the energy interval considered (keV) in the flat region of the spectrum and $\Delta R = R_2-R_1$, where $R_2$ ($R_1$) is the amount of
the bulk  material (cm) necessary to degrade the energy of the alpha from $T_{\alpha}$ to $E_2$ ($E_1$).  We compare our calculation to a rate measurement of alphas from
$^{147}$Sm, ($15.32\,\pm\,0.03$\% of Sm($nat$) and half life of $(1.06\,\pm\,0.01)\times\,10^{11} \text{\,yr}$\,\cite{SamariumLifetime}), and find good agreement, with the ratio between prediction to measurement of $100.2\%\pm 1.6\%\,\text{(stat)}\pm 2.1\%\,\text{(sys)}$. We derive the condition for the flat spectrum, and also calculate the relationship between the decay rate measured at the surface for a [near] surface contamination with an exponential dependence on depth and an a second case of an alpha source with a thin overcoat.  While there is excellent agreement between our implementation of the sophisticated Monte Carlo program SRIM \cite{SRIM} and our intuitive model in all cases, both fail to describe the measured energy distribution of a $^{148}$Gd alpha source with a thin ($\sim200\mu$g/cm$^2$) Au overcoat.  We discuss possible origins of the disagreement and suggest avenues for future study.
\end{abstract}
\maketitle


\vspace{0.2in}

Detectors for neutrinoless double $\beta$ decay ($0\nu\beta\beta$), dark matter and other processes at lower energies ($\lesssim$10\,MeV) can suffer from backgrounds arising from 
alpha decays where the energy of the alpha is degraded from its maximal value of $T_{\alpha}$ by passage through inert material and its total energy is not fully detected in an active detector. Consequently, not all such background events can be eliminated via exclusion cuts around the maximal alpha energy  

Generally, the observed energy of an alpha can be degraded via one of two different scenarios:  either the alpha is emitted from an inert material and is captured in an active detector or, conversely, the alpha exits the active part of a detector and is absorbed in inactive material. For a detector consisting of multiple, discrete elements the total energy of alphas which leave the active volume of one unit  and directly enters the active volume of another unit can typically be reconstructed depending on the detector technology and good knowledge of the relative energy calibration of the detector units.
Given the short absorption length of alphas in matter (typically 10's\,$\mu$m) the alpha emitter in either scenario must lie near the surface of the materials used to construct the detector to result in only partial energy absorption in an active volume.

\begin{figure}[htbp]
\begin{center}
\includegraphics[scale=0.3]{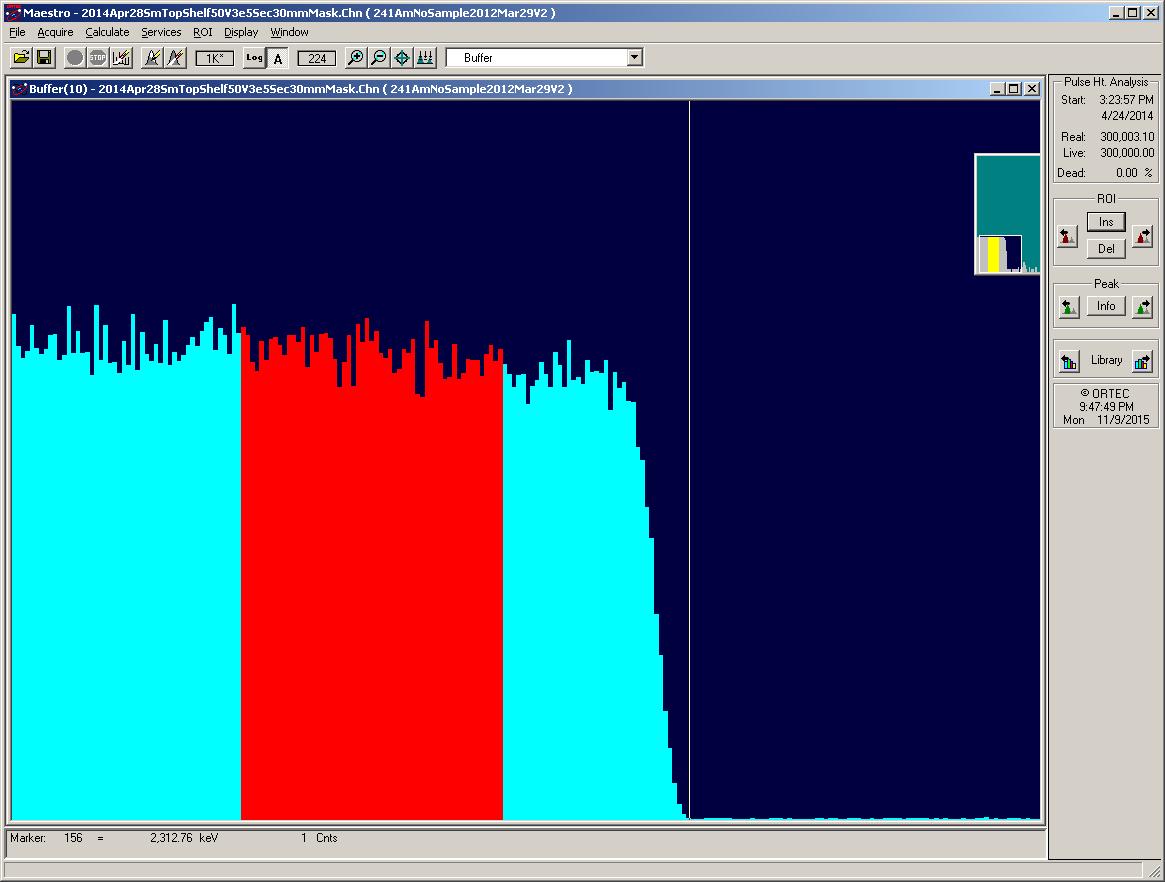}

\caption{Realtime alpha energy distribution from a piece of Sm metal foil. The (red) highlighted area between $1447\,\text{keV}\lessapprox E \lessapprox 1947\,\text{keV}$ is well within the plateau of the decays
which begin below $T_{\alpha}\sim 2310$\,keV, roughly at the energy of the vertical, white line.  The highlighted region contains $\sim$ 44K events in this particular run of $3\times10^5$\,sec. 
The task at hand is to calculate 
the number of decays in the highlighted region using only the known properties of samarium\label{fig:SmAlphaSpectrum}.}
\end{center}
\end{figure}

Figure \ref{fig:SmAlphaSpectrum} is an example of a spectrum of alpha particles from a sample of  99\% pure Samarium metal\,\cite{Goodfellow} (the experimental details are described below). The high end cutoff of the spectrum is at $T_{\alpha}=2310$\,keV\,\cite{TableOfIsotopes} and has a long, flat plateau extending to lower energy. In this figure the hypothetical upper ($E_2$) and lower ($E_1$) energy limits defining a region of the flat plateau are indicated by the red highlighted region.
The task at hand is to calculate the number of events between $E_2$ and  $E_1$ given the known half-life and concentration of $^{147}$Sm in the sample and ultimately the source-detector geometry and measured efficiencies.

\begin{figure}[htbp]
\begin{center}
\includegraphics[scale=0.4]{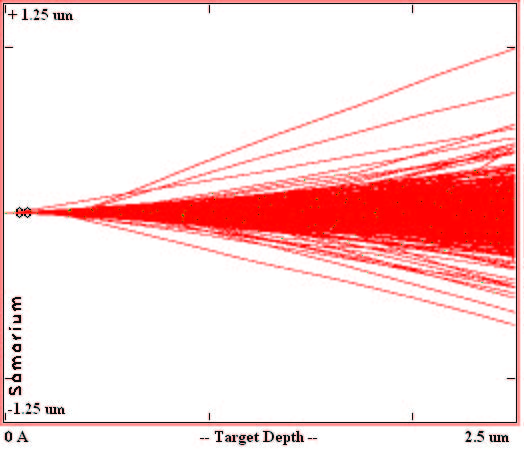}

\caption{Graphics from SRIM of one thousand 2310\,keV alphas transversing a layer of samarium 2.5\,$\mu$m thick. Each line is the trajectory of a single
alpha originating from the origin.}
\label{fig:AlphaTrajectoryA}
\end{center}
\end{figure}

\begin{figure}[htbp]
\begin{center}
\includegraphics[scale=0.75]{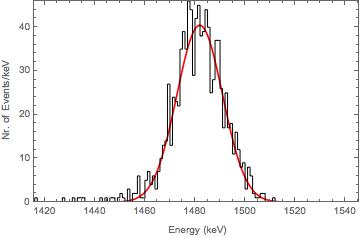}

\caption{Distribution in energy of transmitted 2310 keV alphas incident on 2.5\,$\mu$m thick samarium as generated by SRIM. The histogram (black) is the MC data, and the smooth line (red) is a fit
of a Gaussian distribution\,\cite{Mathematica} with mean $1481.9\pm 0.3$\,keV and $ \sigma $  of $9.1\pm 0.23$\,keV.
}
\label{fig:AlphaTrajectoryB}
\end{center}
\end{figure}

Figure \ref{fig:AlphaTrajectoryA} presents a SRIM\,\cite{SRIM} simulation of alpha particles with an energy 2301\,keV in Sm, at a depth of 2.5\,$\mu$m initially produced in a direction perpendicular to the surface of the samarium. We note that the paths taken by the ions are tightly bundled around their original trajectory. Figure \ref{fig:AlphaTrajectoryB} shows the distribution of the \emph{transmitted} (or remaining) energy of the alpha particles after exiting the surface. The transmitted energy is a narrow distribution (\emph{rms}/mean =0.62\%). Given these two observations, the calculation below makes the simplifying assumptions that the alpha trajectories  from the same location and initial direction follow a common path (\emph{i.e.} no scattering) and their \emph{remaining} energy spectrum has zero spread; \emph{i.e.}  the `spill-out' from one region of phase space will compensate for the `spill-in' from a nearby point in phase space, and vice-versa. 

\begin{figure}[htbp]
\begin{center}
\vskip -.15in
\includegraphics[scale=0.5]{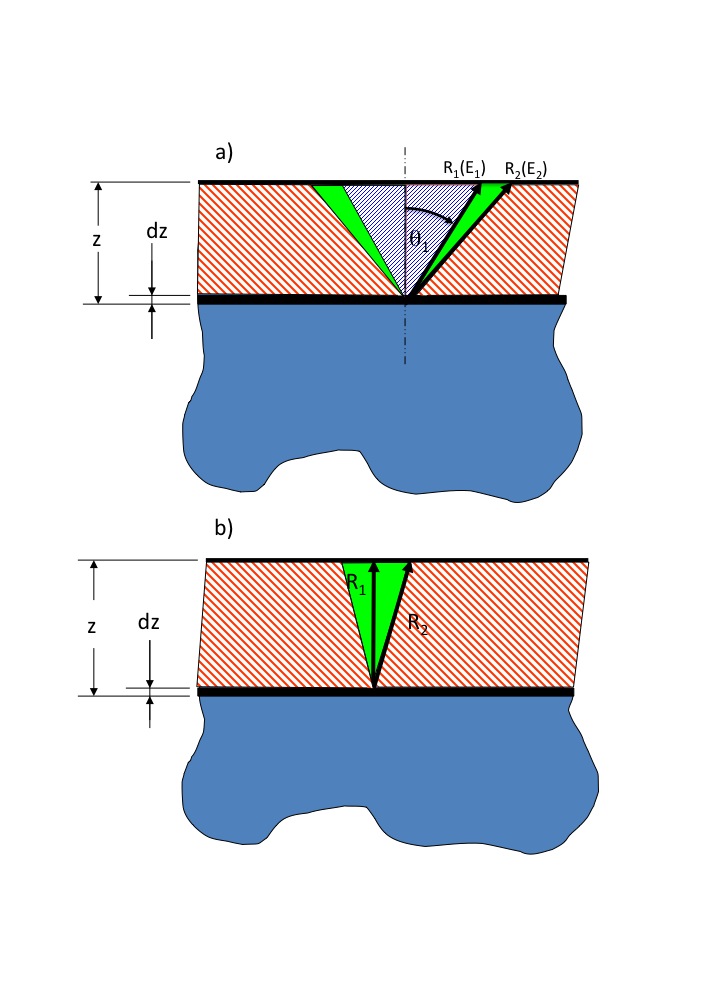}

\vskip -.75in
\caption{Definition of geometry and variable names used in calculations. Only alphas exiting the material between $R_1$ and $R_2$  in the solid green areas will have the correct energy. In the upper panel a) alphas exiting alphas closer to the vertical than $R_1\ (\theta_1)$ (blue, narrow crosshatched area) will retain too much energy and alphas exiting at angles larger than $R_2\ (\theta_2, \text{not shown})$ (red, broad crosshatched area) will have too little energy. In the lower panel b) all alphas from the volume element $dz$ are accepted. In both panels the acceptance volume is cylindrically symmetric around the vertical axis. }

\label{fig:GeometryDefinitions}
\end{center}
\end{figure}

Figure \ref{fig:GeometryDefinitions}-a defines the relationship between $E_1$ ($E_2$) and  the distance  variable $R_1$ ($R_2$), introduces the track angle $\theta$ with respect to the vertical axis, and the depth variable $z$. An alpha particle will on average an
retain an energy $E_1$ ($E_2$) after  traversing a distance $R_1$ ($R_2$). Given a particular depth $z$ there is only a small range in $\theta$ (solid, green shaded) where alphas an satisfy the constraint that they exit the material with energy between $E_1$ and $E_2$.  By the definitions of $R_1$ and $R_2$ and the two assumptions above,  alphas that exit closer to the vertical axis than $\theta_1$ will have too much remaining energy, and alphas which exit with a larger angle of $\theta_2$ will have too little energy.

A volume element of the alpha emitter is $dV = Adz$, where $A$ is its area opposite the detector and $z$ the depth into the material. The formalism below normalizes to unit area, so the area $A$ of the sample cancels out, and $dz$ becomes the surrogate for the volume element $dV$. Invoking azimuthal symmetry around the vertical axis ($\theta_i =0$), the following equalities and the expected decay  rate measured at the surface (Rate$_A$) follow from Fig.\,\ref{fig:GeometryDefinitions}-a for case $A$ where $0 < z < R_1$  :
\begin{eqnarray}
\cos\theta_i & = & {{z}\over{R_i}}, (i=1,2) \\
\text{fractional solid angle} &=&{{2\pi}\over{4\pi}}[(1-\cos\theta_2)-(1-\cos\theta_1)]\\
&\approx& z(R_2 -R_1)/2R_2R_1\\
\text{Rate$_A$}& = &{ {\rho_b(R_2-R_1)}\over {2 R_1R_2 \Delta E}} \int_{0 }^{R_1} z dz \\
 & = & { {\rho_b R_1 \Delta R }\over{4R_2\Delta E} } \text{(\,Bq\,cm}^{-2} \text{\,keV}^{-1} \label{eqn:caseA})
\end{eqnarray}
 As seen in Fig.\,\ref{fig:GeometryDefinitions}-b, (case $B$) a small part of the solid angle is missing 
in this calculation, corresponding to the interval $R_1 < z <R_2$. This can be integrated to yield:
\begin{equation}
\text{Rate}_B = {{\rho_b(\Delta R)^2}\over{4R_2\Delta E}}
\label{eqn:caseB}.
\end{equation}

   Combining Eqn's. (\ref{eqn:caseA}) and (\ref{eqn:caseB}) yields the total rate from the bulk $r_b$:
\begin{eqnarray} 
\text{Total Rate       } \equiv  r_b & = & \text{Rate}_A + \text{Rate}_B \\ 
      r_b  &=  & {{ \rho_b \Delta R}\over{4 \Delta E}}\text{(\,Bq cm}^{-2} \text{\,keV}^{-1}).\label{eqn:FinalExactResult}
     \label{eqn:RateForBulkContamination}
   \end{eqnarray}

 In these relations, $r_b$ is the only measured quantity and $\Delta E/\Delta R$ is obtained from published tables\,\cite{AlphaStoppingPower} or via simple simulation, as is done here using SRIM.  Below we compare our measured surface rate, $r_b$, with the result computed above. We note that $\Delta E/\Delta R$ is not equal to $dE/dx$:    $\Delta E$ refers to a difference in transmitted energy, while $dE/dx$ refers to energy absorbed.


To verify the validity of Eqn.\,(\ref{eqn:FinalExactResult}) we have used a sample of natural samarium metal ($\geq$ 99\% pure, 0.25\,mm thick, with an area of $\sim50  \times 50\text{\,mm}^2$). $^{147}$Samarium decays 100\% via alpha emission, has a natural, average abundance of ($15.0\ \pm\ 0.2)$\%\,\cite{HandbookOfChemistry}, and a half-life of $(1.06\,\pm\,0.01)\times10^{11}$\,yr\,\cite{SamariumLifetime}. It is therefore excellent test candidate for this measurement, having both high abundance (high rate) and essentially flat rate dependence over geologic, if not cosmological time scales.  
 
The detector consisted of NIM\,\cite{NIM} style ORTEC Alpha Aria spectrometer\,\cite{AlphaAria} equipped with a $\sim40$\,mm diameter surface barrier Si detector and readout via a Dell laptop\,\cite{Dell} running the ORTEC Masetro software package\,\cite{AlphaAria}. Because some data acquisition times were long, the full system including the NIM crate, vacuum pump and laptop were powered through an uninterruptible power supply\,\cite{UPS} with enough stored energy to run the system for a few hours in the event of a power interruption.

The energy scale of the system was calibrated by using a mixed $^{148}$Gd\,-$^{241}\text{Am}$ alpha standard source\,\cite{AlphaEnergyCalibration} electroplated on the surface of a 25\,mm diameter stainless steel disk with a reported active area of diameter 17\,mm with no overcoat. The energy scale below $T_{\alpha}(^{148}$Gd) was linearly extrapolated from alpha peaks of these two isotopes. To understand the geometrical acceptance of the detector system a second, recently recently calibrated $^{148}$Gd\,-$^{241}$Am source with a Au overcoat\,\cite{AlphaRateCalibration} was used, corrected for the half-lives of $^{148}$Ge and $^{241}$Am as quoted in \cite{AlphaRateCalibration}, consistent with the values listed in \cite{TableOfIsotopes}. The holder for this source has an outer diameter of 2.54 cm and a reported active area of 5\,mm diameter, with $\sim$200\,$\mu$g/cm$^2$ gold overcoat. Because of the gold overcoat, this second source was not used for energy calibration, but only efficiency measurements.

\begin{figure}[htbp]
\begin{center}

\includegraphics[width=4in]{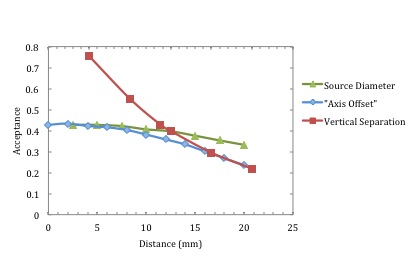}

\caption{Ray tracing Monte Carlo acceptance plotted versus three different spatial parameters:  1) the separation between the source and the detector along a common axis, 2) the diameter of the source at a fixed separation of 13.7\,mm with a common axis, and 3) the offset between the axis of the source and the axis of the detector at a fixed separation of 13.7\,mm. A total of 27 different combinations were examined. }
\label{fig:MCAccptVsDistance}
\end{center}
\end{figure}

\begin{figure}[htbp]
\begin{center}
\includegraphics[width=4in]{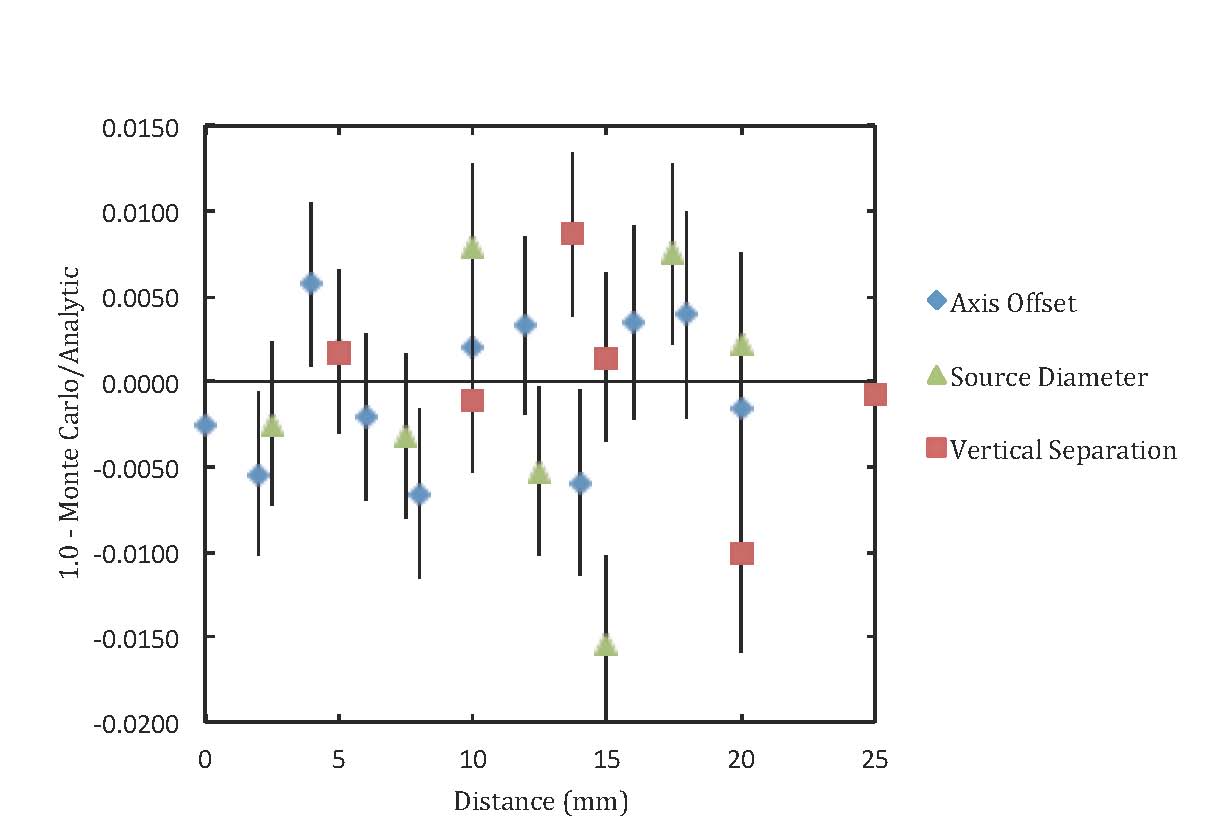}

\caption{ This plot shows measures the reduced ratio (1.0-(Ray Tracing Monte Carlo)/(Analytic Calculation)) for 27 different values of parameters shown in  of Fig.\,\ref{fig:MCAccptVsDistance}. The mean of all the points is $-0.0007$, and the $rms$ about the mean is $0.0056$.}
\label{fig:RatioMCAcceptAnalyticVsDistance}
\end{center}
\end{figure}

A shelf with a cylindrical depression ~25.4\,mm in diameter centered the sources below the Si detector, with a typical source-detector separation in the range 15-17\,mm, measured to an accuracy of 0.1\,mm. A simple, ray tracing Monte Carlo (MC) program computed the geometrical efficiency using the known geometry: a)  assumed straight line trajectories (no scattering) and b) 100\% absorption on any mask or other intervening materials between the source and the detector. For emitters with flat spatial distributions, the Monte Carlo calculation was compared with an analytic estimate\,\cite{AnalyticalAcceptance}. The ratio of the MC acceptance computed with 100k events into solid angle $\Omega=2\pi$ to the analytic acceptance for a variety of geometries with vertical separations from 5 to 25\,mm, axis offsets between the source and detector or source diameters, both in the range  0-20\,mm, was 1.0007 with an \emph{rms} for the distribution 0.0056 (a total of 27 different geometrical configurations were tested). The results are graphically displayed in Figs.\,\ref{fig:MCAccptVsDistance} and \ref{fig:RatioMCAcceptAnalyticVsDistance}. We used the difference between the acceptance computed from the ray tracing MC to the analytic calculation as our systematic error on the acceptance. 

A Telfon mask with a 0.075" (0.19\,mm) hole that could be positioned at intervals across the source along  two perpendicular axes allowed the spatial distribution of the activity to be measured (this was done prior to gold plating). The measurement showed the distribution of both the $^{241}$Am and $^{148}$Gd isotopes to be a roughly Gaussian distribution in shape with  $\sigma 's=$(1.3, 1.5)\,mm\,$\pm$\,0.1\,mm in the two perpendicular directions and average values = (0.2, 1.3)\,mm\,$\pm$\,0.2\,mm along the same axes, respectively, both cutoff beyond a radius of $\gtrsim$3\,mm. The ratio between a MC calculation for a flat, coaxial distribution  for this source and a Gaussian Distribution with an axis offset of 1.4\,mm from the symmetry axis of the detector and $\sigma_r$=1.4\,mm was 1.007 $ \pm$0.007  (stat.) We assign the difference to the systematic error.  

We were concerned about potential non-uniformities in the Si Detector efficiency over its area. Comparison of the rate on centerline with the rates measured at four extreme locations $(\sim \pm 13$\,mm along two perpendicular axes as allowed by the vacuum enclosure) using a teflon collimator  (aperture of  3.2\,mm ID \,$\times\,$0.27 \,mm thick), checks the spatial uniformity of the detector. The ratio of the measured rates at the extrema compared to the rate on centerline was calculated from the analytic acceptance formula, yielding an overall agreement  of $1.003\pm 0.004$ (stat) in the four ratios. We assume assume spatial uniformity is perfect, and apply our measured difference from unity as a systematic error.

To check our results, we compared our rate calculation of the certified, calibrated, gold plated ($\sim$200\,$\mu$g/$cm^2$)  $^{148}$Gd\ $\textendash^{241}$Am mixed source to the results supplied by the vendor. Our results are reported for the individual isotopes and the combined rate in Table \ref{table:AmGdSourceRate}. The vendor provides both statistical and systematic errors at the 99\% confidence level. In Table \ref{table:AmGdSourceRate} we report these errors divided by 2.575 so we could add them in quadrature with errors derived for this work.  We observe that the agreement between our rate measurements and the vendor is quite good, with less than 1\% difference to the vendor values and total errors of less than $\lesssim$2\%. These results provide legitimacy that we should be able to measure the surface rate of $^{147}$Sm with errors on the order of 1-2\%.

\begin{table}[htdp]  
\caption{Calculation of rates and errors from a mixed, rate calibrated  $^{148}$Gd and $^{241}$Am alpha source. A: Common factors for both isotopes; B, C, and D: values and results for 
$^{148}$Gd, $^{241}$Am and their combined rates, respectively. }
\begin{center}
\begin{tabular}{lccc}
\multicolumn{4}{c} {A:  Common Values and Errors }\\
  & & Statical     & Systematic  \\ 
 Item (units) & Value  & Error    & Error \\ 
 \hline\hline
 Geometric Efficiency  &  0.373    & $\pm0.001  $ & $\pm0.002 $     \\ 
Si Detector Uniformity    & \ \ \ 1.0  &  $\pm0.004$  & $\pm0.003$  \\
(Efficiency)  Error from& \multirow{3}*{\  \  \ (1.0)} &\multirow{3}*{ -} & \multirow{3}*{ $\pm0.007$}\\ 
\ \ Source-Detector  0.1\,mm       &  \\
\ \ Separation Uncertainty  & \\
Grand Total Acc. \& Uniformity & 0.373 & $\pm0.002 $&  $\pm0.007$  \\
Livetime this experiment  (sec) &  1,000 & - & - \\ 
Elapsed time between  & \multirow{3}*{1.281} &\multirow{3}*{ -} &\multirow{3}*{  $\pm0.003$}  \\
\ \ vendor calibration  and &\\
\ \ this measurement  (yr) & \\
\hline \hline
\multicolumn{4}{c} { }\\
\multicolumn{4}{c}{ B:   $^{148}$Gd}    \\
  & & Statical     & Systematic \\ 
 Item (units)& Value  & Error    & Error  \\ 
 \hline\hline
  Vendor Rate into  2$\pi${\it sr} (Bq) & 375.9 &    $\pm1.9 $ &  $\pm4.4$ \\  
Lifetime (yr) & 108.2 & - &  $\pm4.3$   \\ \hline
Corrected Vendor Rate (Bq) & 371.4   & $\pm1.9$ &$\pm4.4$  \\ 
Raw Rate, this exp. (Bq)  &143.7 &  $\pm0.4$  & - \\
Background (Bq)  & \ \ \ 4.0  &$\pm0.3$&$\pm0.4$  \\  
Net Rate into $2\pi${\it sr} (Bq)& 374.7 &$\pm3.4$ &$\pm7.4$ \\
Ratio This Exp./ Vendor&1.009& $\pm0.011$ &$\pm0.023$  \\  
\hline\hline
\multicolumn{4}{c} { }\\
  \multicolumn{4}{c}{ C:  $^{241}$Am}  \\
   & & Statical     & Systematic \\ 
 Item (units) & Value  & Error    & Error \\  
 \hline\hline
Vendor Rate into 2$\pi${\it sr} (Bq) & 379.5  & $\pm1.9$& $\pm4.4$ \\
Lifetime (yr) & 623.5 & -& $\pm1.0$ \\ \hline
Corrected Vendor Rate (Bq) & 378.6 & $\pm1.9 $& $\pm4.4$ \\
Raw Rate this exp. (Bq) & 143.4 & $\pm0.4$ &- \\
Background (Bq)&\ \ \ 4.0 &  $\pm0.3$ &- \\ 
Net Rate into $2\pi${\it sr} (Bq) & 374.0 &  $\pm3.4$&$\pm7.9$   \\
Ratio This Exp./Vendor & 0.988 & $\pm0.011$  & $\pm0.022$ \\  \hline\hline
\multicolumn{4}{c} { }\\
\multicolumn{4}{c}{ D:  Combined rates  $^{148}$Gd  + $^{241}$Am}  \\
   & & Statical     & Systematic \\ 
 Item (units)& Value  & Error    & Error \\  
  \hline\hline
Total Vendor Rate Am+Gd (Bq) & 750.1 & $\pm2.7$  & $\pm6.2 $  \\
Total Rate This Exp. Am+Gd (Bq) & 748.7 &$\pm4.8 $ & $\pm10.3$  \\
Ratio  This Exp./Vendor & 0.998 & $\pm0.007 $ & $\pm0.016$ \\
\hline\hline
\end{tabular}
\end{center}
\label{table:AmGdSourceRate}
\end{table}%

We now proceed to our measurement of the surface decay rate for $^{147}$Sm.
The nominal world average, fractional abundance of $^{147}$Sm in Samarium metal is 0.1499 \,\cite{HandbookOfChemistry}. We obtained a sample of 
Sm metal foil 5\,cm $\times$5\,cm $\times$0.25\,mm with a reported purity of  99\%\,\cite{Goodfellow} or better. The Sm metal was kept under vacuum when not in use, but some white oxide appeared over year the data was gathered (the lifetime of Sm oxide is measured to be the same as Sm metal\,\cite{SamariumLifetime}). The isotopic concentration of our sample was measured using Inductively Couples Plasma Mass Spectrometry (ICPMS)\,\cite{147SmIsotopeFraction} using 15 replicate analyses to compute the mean isotopic concentrations and uncertainties.  The $^{147}$Sm isotopic abundance in our sample was 0.1532 $\pm 0.0006 \text{(stat)}$ (95\% Confidence level) or about 2\% higher than the world average. We assign a systemic error of 1.0\%, consistent with the reported sample purity.

The lifetime of Sm is $4.83\times10^{18}$\,sec, {\it i.e.} longer than the lifetime of the universe. Hence, there is no need to correct the $^{147}$Sm concentration
for decay processes. We use the commonly published value for the density\,\cite{HandbookOfChemistry}.

A 0.250\,mm thick teflon mask, with transverse dimensions slightly larger than the Sm foil and with a circular apature $30.64 \pm\,$0.10\,mm $rms$ centered below the Si alpha detector defines the active area of the Sm for this measurement. 
In Table~\ref{table:SmBulkAndSurfaceRate} we present the calculated rates for the bulk decay rate, surface decay rate  and our measurement of the latter. The data reported 
here represent the sum of four different runs with differing livetimes and geometrical acceptances. We add the data and report the livetime weighted geometrical acceptance and its error.
We calculate the bulk decay rate is $954.9\pm 1.9 \text{(stat)} \pm 1.6 \text{(sys) Bq/keV-cm}^3$, and our intuitive model described above predicts the decay rate at the surface to be $6.82 \pm 0.02\text{\,(stat)}\pm0.01\text\,(sys) \times 10^{\textendash 5} \text {\,Bq/keV/cm}^2$. We measure a rate of $6.60 \pm 0.01\text{\,(stat)}\pm0.10\text{\,(sys)} \times 10^{\textendash 5} \text {\,Bq/keV/cm}^2$, or a ratio of expectation to this experiment of 102.3\% $\pm1.6\%\text{\,(stat)},\pm2.1\%\text{\,(sys)}$. Hence, we see that our model for the surface decay rate yields results in excellent agreement with the measured rate and the scale of the errors are consistent with those derived above for the rate of the calibrated, mixed $^{148}$Gd\,\textendash $^{241}$Am alpha source.

\begin{table}[htdp]
\caption{ A: Calculation of the bulk decay rate ($\rho_b$) of $^{147}$Sm from known parameters with errors.  B: Parameters values and estimated surface rate above pure $^{147}$Sm material
for a particular energy interval.}
\begin{center}
\begin{tabular}{lcccc}

\multicolumn{5}{c} { A: Parameter values  to calculate {\it predicted}  bulk decay rate}\\ 
                         &                        & Statistical   & Systematic   \\
   Item (units)                                               &      Value                                    &  Error        &      Error               \\
 \hline \hline
$^{147}$Sm Lifetime$^{\star} $   (sec)          &  $4.83\times10^{18}$           &       -              &   $0.05 \times 10^{18}$   \\ 
Density (g/cm$^3$) & 7.520                             &                -               &            -                               \\                        
Atomic Weight (g) \cite{SamariumAtomicWeight}        & 150.36                         &  -                          &    0.02         &       \\ 
$^{147}$Sm abundance &  0.1532                    & 0.0003               &   0.0002      &             \\
Purity  & 0.99 &  - & 0.01 \\
 \hline 
Bulk Decay rate &   \multirow{2}[3]*{   $945.3 $}   & \multirow{2}*{$\pm 1.9 $ \bigstrut} & \multirow{2}*{$\pm 9.6 $\bigstrut}                 \\ 
 \ \ $\rho_b$ (Bq/cm$^3$)   &                 \\  
\hline \hline
\multicolumn{5}{ c }{      } \\
\multicolumn{5}{c}  {B:  Parameter values to calculate {\it predicted} surface decay rate} \\ 

                          &                        & Statistical   & Systematic  \\ 
   Item (units)                                              &          Value                                &  Error        &      Error              \\ \hline \hline

Energy Interval                      &  \multirow{2}*{ 507.4}                            &  \multirow{2}*{  0.8}        & \multirow{2}*{  -}            \\
\ \  SRIM $\Delta E$ (keV) &          \\              
SRIM Distance $\Delta R$ (cm)   & $1.45\times 10^{-4}$ &-          &       -                 \\ 

\hline 
Surface Rate  $r_b$ &   \multirow{2}*{ $6.75$ }& \multirow{2}*{ $\pm 0.02 $ } & \multirow{2}*{$\pm 0.01 $ } \\
\ \  (Bq/keV-cm$^2 \times 10^{-5}$) &                            \\
\hline \hline

\multicolumn{5}{ c }{      } \\
\multicolumn{5}{c}  {C:  Parameter values to calculate {\it measured} surface decay rate} \\ 
                          &                         & Statistical   & Systematic   \\
  Item (units)                                               &    Value                                     &  Error        &      Error              \\
\hline\hline                                               
Mask area ($cm^2$)   &        7.354                        &    -                 &    0.05      \\
Energy Interval                      &  \multirow{2}*{ 499.5}                            &  \multirow{2}*{  3.7}        & \multirow{2}*{2.9}                \\
\ \  $\Delta E$ (keV)  &          \\              
Livetime weighted & \multirow{2}*{0.4932}   & \multirow{2}*{ 0.0037}  &  \multirow{2}*{0.0026}      \\
\ \ Geometric Efficiency  & \\
Livetime  (sec) & 587,369&  -   & $< 0.02$   \\
Events in $\Delta E$ & 70,959 & 267 & - \\
Background                 & 85 & 9 &  -  \\
Events - Background  & 7,773  & 267  & -  \\  \hline \hline
Measured Surface Rate & \multirow{2}* {$6.60 $} & \multirow{2}*{$\pm0.008$} & \multirow{2}*{$\pm 0.10$} \\
\ \  (Bq/keV-cm$^{2} \times 10^{\textendash 5} $) & \\ \hline \hline
Ratio & \multirow{2}*{ 102.3 } & \multirow{2}*{  $\pm 1.6$ }  &  \multirow{2}*{$\pm 2.1 $} \\ 
\ \ Prediction/This Exp. (\%)  &   \\ \hline \hline
\multicolumn{5}{l} {$\star$ Lifetime is presented for completeness, but not used in calculation, } \\
\multicolumn{5}{l} {since the lifetime exceeds the age of the universe.}
\end{tabular}
\end{center}
\label{table:SmBulkAndSurfaceRate}
\end{table}

Recall that in Eqn.\,(\ref{eqn:FinalExactResult}) that $\Delta E/\Delta R \neq dE/dx$. The $\Delta E$ referred to is the difference in the {\it transmitted}
(or observable) energies, while $dE/dx$ measures the energy lost.
Importantly, we can now understand the requirement for a flat distribution:  if the energy dependence of $dE/dx$ in the region of interest is the  form ${{dE}\over{dx}}(E) = a + b E$, where a and b are constants, and any quadratic or higher terms are
negligible, then the rate of alpha decay measured at the surface will be a flat plateau independent of the transmitted energy for a material with a uniform bulk contamination.

The same techniques used to calculate the surface rate in Samarium with a uniform density of $^{147}$Sm can be extended to the case of a material with a distribution
of emitters exponentially distributed into the material. This might be the case, for example, for environmental contamination deposited on the surface with the radioactive daughers recoiling into the material followed by diffusion into the substrate \cite{CUORERadonRome}.  To compute the rate in this case, we need only replace the bulk density in cases A, Eqn.\,(\ref{eqn:caseA}), and B, Eqn.\,(\ref{eqn:caseB}) above, 
by $\rho_b \rightarrow \rho_{\lambda} e^{-z/\lambda}/\lambda$, where $\lambda$ describes the exponential distribution of the daughter into the material as a function of the depth, $z$, and $\rho_{\lambda}$ is a constant equal to the density of the isotope at the surface. In this case we find the surface rate $r_{\lambda}$\ is: 
\begin{equation}
r_{\lambda}={\rho_{\lambda}{\lambda^2} \over{ 2 \Delta E R_1 R_2}}
[ R_1( e^{-(R_2/\lambda)} -1) + R_2(1 - e^{-R_1/\lambda})]
\end{equation}

Analogous to Eqn.\,(\ref{eqn:RateForBulkContamination})  $   {\underset{\lambda\to\infty}\lim}
r_{\lambda}\rightarrow \rho_{\lambda}\Delta R/4 \Delta E$,  as expected for a constant density of an alpha emitters, and ${\underset{\lambda\to 0}\lim}\  r_{\lambda}\ =0$, since the total alpha contamination tends towards zero as $\lambda \rightarrow 0$.

\begin{figure}[htbp]
\begin{center}
\includegraphics[width=4in]{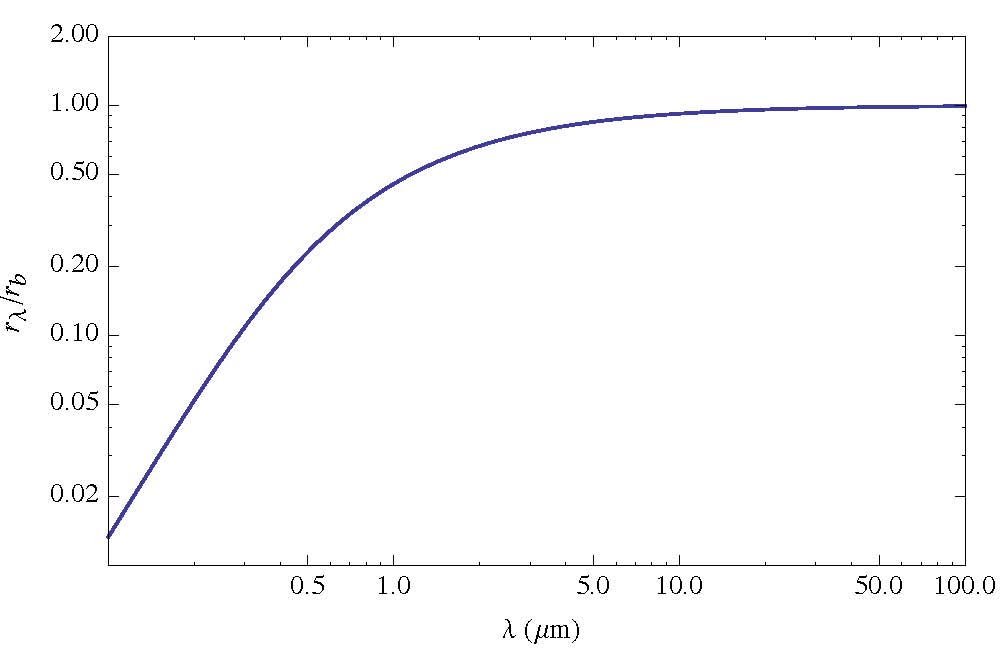}
\caption{Ratio of the rate $r_{\lambda }$ from surface contamination to the rate from bulk contamination $r_b $ plotted as a function of the depth penetration  $\lambda$ of the surface contamination and normalized so that ${\underset{\lambda\to\infty}\lim}   r_{\lambda} \rightarrow r_b$. In this illustration $R_1 =1.0\,\mu$m, $R_2=1.5\,\mu$m; the 
energy range $\Delta E$  and other factors cancel in the ratio $\rho_{\lambda} /\rho_b$.}
\label{fig:AlfaSurfaceEffectVsLambda}
\end{center}
\end{figure}

\begin{figure}[htbp]
\begin{center}
\includegraphics[width=4in]{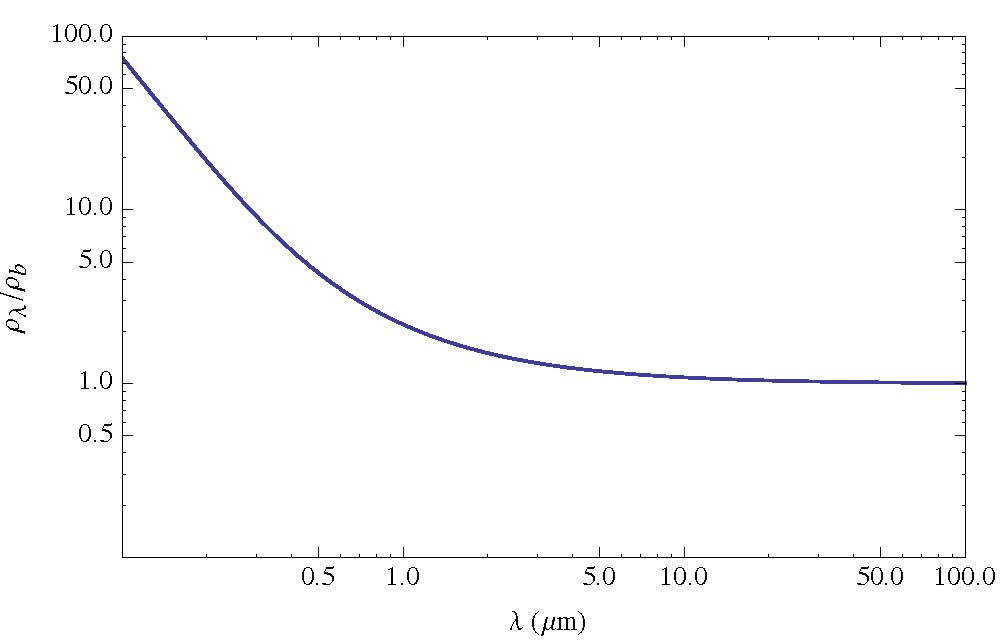}
\caption{Ratio of $\rho_{\lambda}/\rho_b$ versus  the exponential depth penetration  $\lambda$ needed to  maintain the same surface rate as would follow from a  bulk contamination of $\rho_b.$
 Any point along the line will satisfy the condition that $r_{\lambda}=r_b$. For this illustration $R_1 =1.0\,\mu$m, $R_2=1.5\,\mu$m; the 
energy range $\Delta E$  and other factors cancel in the ratio $r_{\lambda} /r_b$.}
\label{fig:ConstantAlphaSpectrumVsLambda}
\end{center}
\end{figure}

For a given rate measured at the surface, there is an ambiguity between the exponential depth distribution of emitters ({\it i.e. $\lambda$} ) and the surface density ($\rho_{\lambda}$) that can account for the measured rate. We illustrate this via two examples: in Fig.~\ref{fig:AlfaSurfaceEffectVsLambda} we show the ratio of the surface rate to the rate for constant  bulk contamination 
as a function of the depth penetration; and in Fig.~\,\ref{fig:ConstantAlphaSpectrumVsLambda} we plot the value of $\rho_{\lambda}$ versus  the exponential depth penetration  $\lambda$ needed to  maintain the same surface rate that would follow from a constant bulk contamination of $\rho_b.$

The energy distribution of transmitted alphas can resolve the ambiguity between their exponential depth distribution and their density. In our model it is
easy to write down the energy distribution, $N_{\lambda}(E)$, of the transmitted alphas:
\begin{equation}
N_{\lambda}(E) = \int ^{R_{max}}_0 dz \int^{R_{max}}_z dR \left( {{e^{-z/\lambda} }\over{\lambda}} \right) \left({{z}\over{R^2}}\right) {{e^{-(E-(T_{\alpha}- R{{dE}\over{dx}} ))^2/2\sigma^2}\over{\sqrt{2\pi}\sigma}}}.
\label{eqn:ExponentialEnergyDist}
\end{equation}

In Eqn.\,(\ref{eqn:ExponentialEnergyDist}) the first term inside the integral is the exponential distribution of the alpha emitters versus the depth $z$. The second term is proportional to the solid angle, and the third term represents the distribution of the transmitted energy degraded from its initial energy $T_{\alpha}$ by an amount $R\cdot dE/dx$ and smeared by a resolution $\sigma$. Our intuitive model does not predict the intrinsic energy spread $\sigma$ of the transmitted alphas, which is normally of order $\lesssim 10$\,keV ({\it{e.g.}} the width of the Gaussian distribution  of the alpha spectrum in Fig.\,\ref{fig:AlphaTrajectoryB} for a perfect detector). However, the resolution of our detector is about 30\,keV so the contribution of the intrinsic energy spread is small. The limit of integration $R_{max}$ is the maximum distance that an alpha can travel in the material. In our approximation, $R_{max} =T_{\alpha}/(dE/dx )$, or it can be one parameter in a fit to the energy distribution. It is worth noting that the limits of the second integral depends on the first:  the distance of the material transversed by the alpha cannot be smaller than the vertical depth of its origin. 

It is possible to reverse the order of the integration, and perform the integral over $z$ first, but care must be taken with the limits of integration. Also, some of the terms loose their simple interpretation:
\begin{equation}
N_{\lambda}(E) =  {{1}\over{\sqrt{2\pi}\sigma}}\int ^{R_{max}}_0 dR {{(\lambda-e^{-R/\lambda}(\lambda+R))e^{-(E-(T_{\alpha}- R{{dE}\over{dx}} ))^2/2\sigma^2}}\over{R^2}}.
\label{eqn:ExpDistAfterIntegrationOverZ}
\end{equation}
There is an apparent pole in this integral at $R\rightarrow0$, but expanding the first term from the integral over $z$ we find it exactly cancels the $1/R^2$ behavior.
We integrate this expression numerically\,\cite{Mathematica}.

One should note that in this intuitive estimation for the energy distribution, there are, in principal, no free parameters; when fitting to data or MC distributions there is an  overall normalization constant $(A)$ fixed to: 
\begin{equation}
 A = \text {(number of events)}\times\text{(histogram bin width)}/\int N(E)dE,
 \label{eqn:NormalizationConstant}
 \end{equation}
where the limits of the integral over energy are picked to correspond to the dataset range. Indeed, as we show below, there is spectacular agreement with SRIM just using the naive constants; 
5-parameter fits ($ \lambda,\ T_{\alpha},\ dE/dx,\ \sigma,$ and the normalization $A$) versus energy ($E$) typically yield $\chi^2$/DOF of order 1 or better \,\cite{Mathematica}.

\begin{figure}[htbp]
\begin{center}
\includegraphics[scale=0.5]{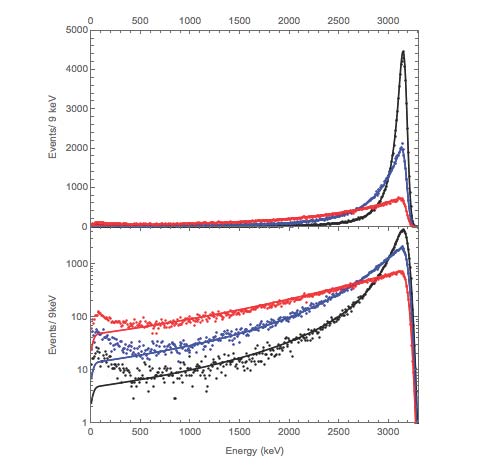}
\caption{Comparison between SRIM (data points) and Eqn.\,(\ref{eqn:ExpDistAfterIntegrationOverZ}) for three values of $\lambda = 0.1\,\mu$m (black-highest near 3100\,keV),
$\lambda = 0.3\,\mu$m (blue-middle) and $\lambda = 1.0\,\mu$m (red-lowest) in gold. The parameters of our model are fixed to their naive values, see Table \ref{table:FitsToExpContamination}. The same information is plotted on both linear (upper panel) and logarithmic (lower panel) vertical scales.}
\label{fig:SRIMVsNaiveParametersExpContamination}
\end{center}
\end{figure}

In Fig.\,\ref{fig:SRIMVsNaiveParametersExpContamination} we compare three data sets with $\lambda = 0.1, 0.3$  and  1.0\,$ \mu$m, generated with SRIM and Eqn.\,(\ref{eqn:ExpDistAfterIntegrationOverZ}) using the naive parameters (see Table \ref{table:FitsToExpContamination}) for alpha contamination in gold.  In general there is excellent agreement between the data points and Eqn.\,(\ref{eqn:ExpDistAfterIntegrationOverZ}), except near low energy where the value of $dE/dx(E)$ is rapidly changing and Eqn.\,(\ref{eqn:ExpDistAfterIntegrationOverZ}) underestimates the data. 

\begin{figure}[htbp]
\begin{center}
\includegraphics[scale=0.5]{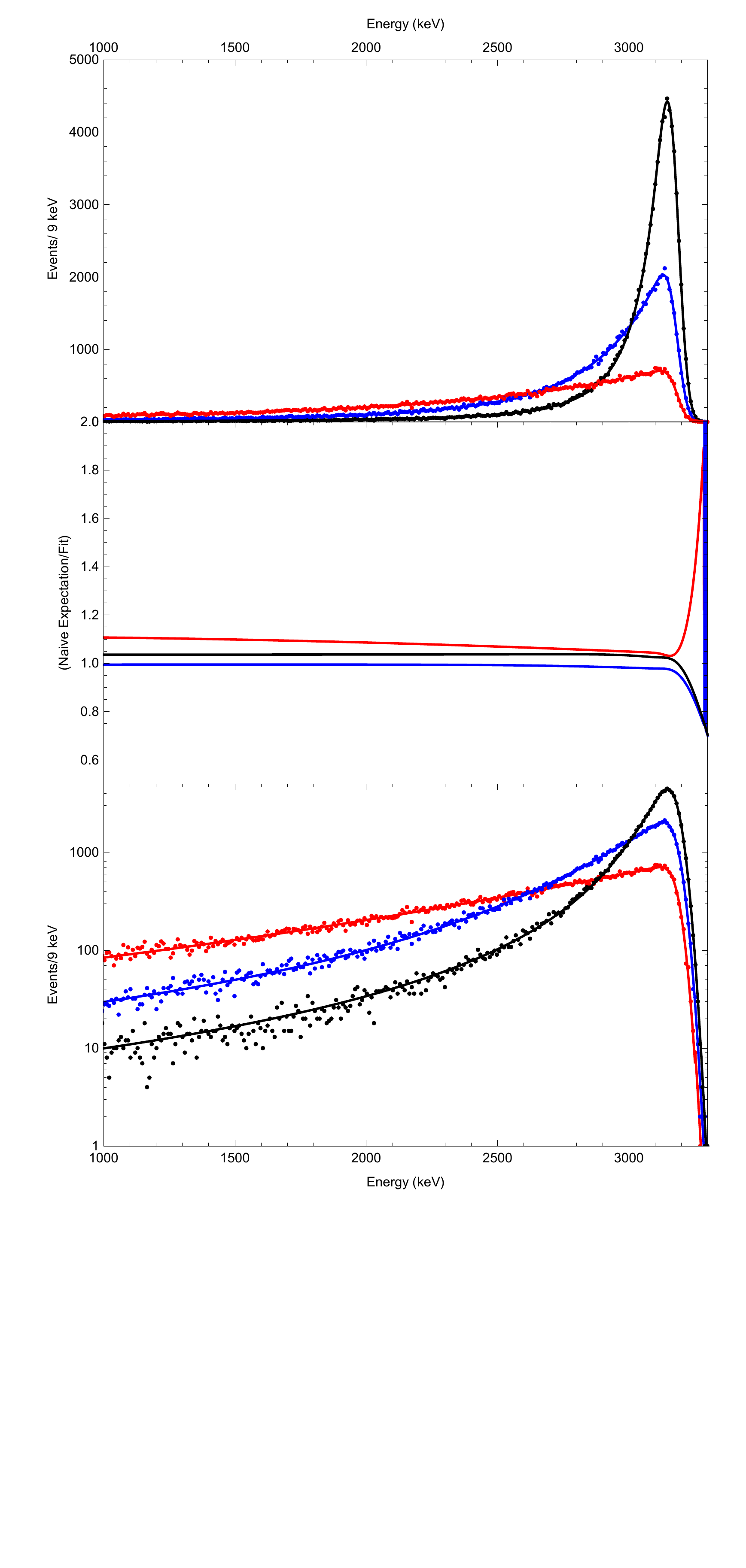}
\caption{Comparison of the SRIM MC data versus five parameter fits using Eqn.\,(\ref{eqn:ExpDistAfterIntegrationOverZ}) normalized to the for three different exponential depth concentrations of alpha emitters in gold. The color scheme is the same as in Fig.\,\ref{fig:SRIMVsNaiveParametersExpContamination}.
The top figure is plotted on a linear vertical scale, the middle figure shows the ratio between Eqn.\,(\ref{eqn:ExpDistAfterIntegrationOverZ})  evaluated with the naive parameters to the fitted parameters (see Table\,\ref{table:FitsToExpContamination}), and the bottom plot is the same information as the uppermost, except on a logarithmic vertical scale. We see the evaluation of the function with naive parameters agrees to a few percent to the function evaluated with the fit parameters, except in the rapidly rising region above $\sim$3150\,keV.  }
\label{fig:SRMandRWKFitLamba-0.1-0.3-1.0}
\end{center}
\end{figure}

In Fig.\,\ref{fig:SRMandRWKFitLamba-0.1-0.3-1.0} we show the SRIM data for $1000\,\text{keV}\leqq E \leqq 3000\,\text{keV}$ in comparison with the 
model fit with five free parameters as explained above and listed in Table \ref{table:FitsToExpContamination}. Additionally, in the center panel, we show the ratio between Eqn.\,(\ref{eqn:ExpDistAfterIntegrationOverZ}) using the naive and fit values of the parameters. We report the $\chi^2/$DOF for the fit in the last column. The fits are 
exceptionally good, with $\chi^2/$DOF $\leq 1.0$. The fit function differs from the naive expectations  mostly in the rapidly change region on the higher energy side of the peak(s), as can
seen in the middle plot of Fig.\,\ref{fig:SRMandRWKFitLamba-0.1-0.3-1.0}
\begin{table}[htdp]
\caption{Results from SRIM MC data for three different exponential depth concentrations of alpha emitters in gold compared to a five parameter fit of Eqn.\,(\ref{eqn:ExpDistAfterIntegrationOverZ})
to the data the energy interval $1000<E<3300$\,keV. The three sets  of values (highest to lowest at 3150\,keV)
are for $\lambda = 0.1, 0.3 \text{ and } 1.0\,\mu$m depth distributions. All the SRIM MC data share the same generated initial energy $T_{\alpha}=3182$\,keV, $dE/dx = 552$\,keV/$\mu$m
and the transmitted energies were smeared with Gaussian distribution  of zero mean and $\sigma=30$\,keV representing the detector resolution. The differing values of the amplitude $A$ for the raw MC data reflect the differing number of the 100K generated alphas that exit the surface. }
\begin{center}
\begin{tabular}{ccccccc}

  $\lambda$  & $A$ &  $T_{\alpha}$  & $dE/dx$ & $\sigma$  &$ \chi^2 $\\
  ($\mu$m) & ($\times10^3$) & (MeV) &(keV/$\mu$m) & (keV) &$\overline{ \text{DOF}}$ \\
\hline\hline
  $\lambda\equiv 0.1$ & 877  & 3182.0 & 552.0  & 30.0  & \\
  $0.1000\pm0.0002$ & $868\pm12$ & $3182.6\pm1.6 $ &$545.1\pm 5.2 $ & $30.5\pm0.3$ & 0.97\\
  \hline
$\lambda \equiv 0.30$ & 829  & 3182.0 & 552.0  & 30  & \\     
$0.282\pm0.005$ &$ 860\pm10$ & $ 3182.7 \pm3.5$ & $581\pm12$ & $30.5\pm 0.5$ & 0.92\\
\hline
$\lambda \equiv 1.0$&  885  & 3182.0 & 552.0  & 30  & \\
$1.003\pm0.001$ & $822\pm4$ & $3184.2\pm1.4$ &$ 531.7\pm4.4$ & $28.7\pm0.8$ &0.89\\ 
\hline\hline
\end{tabular}
\end{center}
\label{table:FitsToExpContamination}
\end{table}%

It is simple to extend this analysis to the energy distribution of an alpha source with a overcoat ({\it eg.}\,gold) of thickness $R_0$, such as described earlier to verify the sensitivity of our apparatus to rate calculations.
Recycling the notation used above, the energy distribution, $N_{oc}(E)$, is given by:

\begin{equation}
N_{oc}(E)=   \int ^{R_{max}}_{R_0} dR    \left( {{R_0}\over{R^2}}\right) \left({{e^{-(E-(T_{\alpha}-R{{dE}\over{dx}}))^2/2\sigma^2}} \over{\sigma\sqrt{2\pi}}}\right).
\label{eqn:OvercoatEnergyDistribution}
\end{equation}
where the first term is proportional to the solid angle, and the second term is the energy $E$ of the transmitted alpha degraded from its  initial energy $T_{\alpha}$ by $R\cdot dE/dx$ from transmission through the overcoat and smeared by the detector resolution, $\sigma$. The limits of integration are such that the amount of material transversed by the alpha can never be less than the thickness of the overcoat. There are typically five fit parameters ( an amplitude $A_{oc},\, R_{max},\ T_{\alpha},\ dE/dx \text{ and } \sigma$). We leave $R_0$ fixed at 
0.1\,mm, as it has little influence on the fit except to slightly modify the shape \emph{above} the peak. In the limit $R_{max}\rightarrow R_0$ we note the formula collapses to a simple Gaussian distribution with mean energy ($T_{\alpha}-R_0\cdot dE/dx$) and resolution $\sigma$.

\begin{figure}[htbp]
\begin{center}
\includegraphics[scale=0.5] {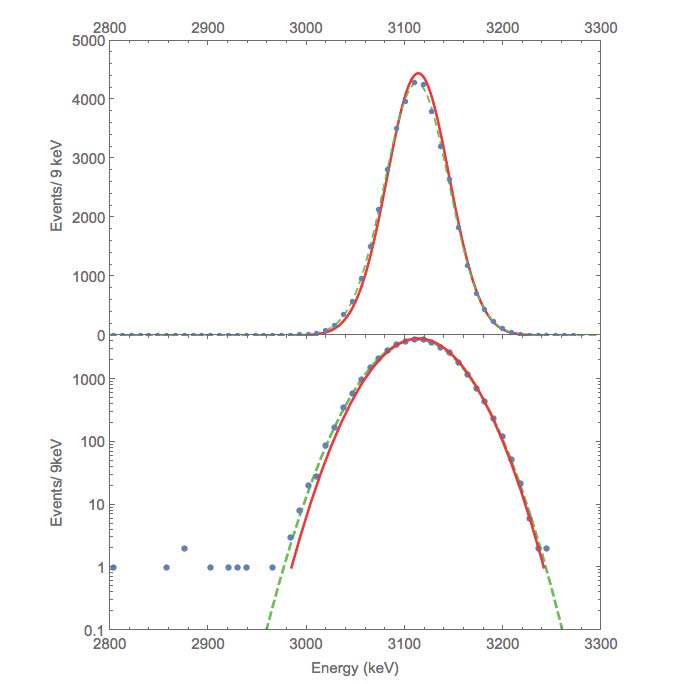}

\caption{Comparison between SRIM MC data (blue points), the analytic expression with naive constants (red, solid line) and the analytic expression fitted to the data (green dashed line) for alphas from a  $^{147}$Gd source ($T_{\alpha}=3182.8$\,keV) with a  $\sim 200\mu$g/cm$^2$ gold overcoat. The same information is plotted on a linear vertical scale (upper panel) and a logarithmic vertical scale (lower panel).}
\label{fig:SRIMComareRWAuOveroat}
\end{center}
\end{figure}

In Fig.\,\ref{fig:SRIMComareRWAuOveroat}  we compare SRIM data with Eqn.\,(\ref{eqn:OvercoatEnergyDistribution}) evaluated with both naive and fitted values, see Table \ref{table:FitsToGoldOvercoat}.  In the naive parameters, $R_{max}$ is listed at 0.16\,$\mu$m corresponding to the nominal Au overcoat thickness of 0.1\,$\mu$m$/\cos 53^{\degree}$, the maximum angle allowed by our source-detector geometry. We observe reasonable agreement between the SRIM MC data and Eqn.\,(\ref{eqn:OvercoatEnergyDistribution}), normalized to the data according to Eqn.\,(\ref{eqn:NormalizationConstant}). Also show is $N_{oc}(E)$ fit to the MC data, showing excellent agreement with a $\chi^2/\text{DOF}$ of 0.49. The errors returned on the parameters are very large, however, indicative the shape is a nearly Gaussian Distribution. There is a low energy tail of a few events in the SRIM MC data, which may, or may not, be important depending on the goal of a particular analysis.

\begin{figure}[htbp]
\begin{center}
\includegraphics[scale=0.5]{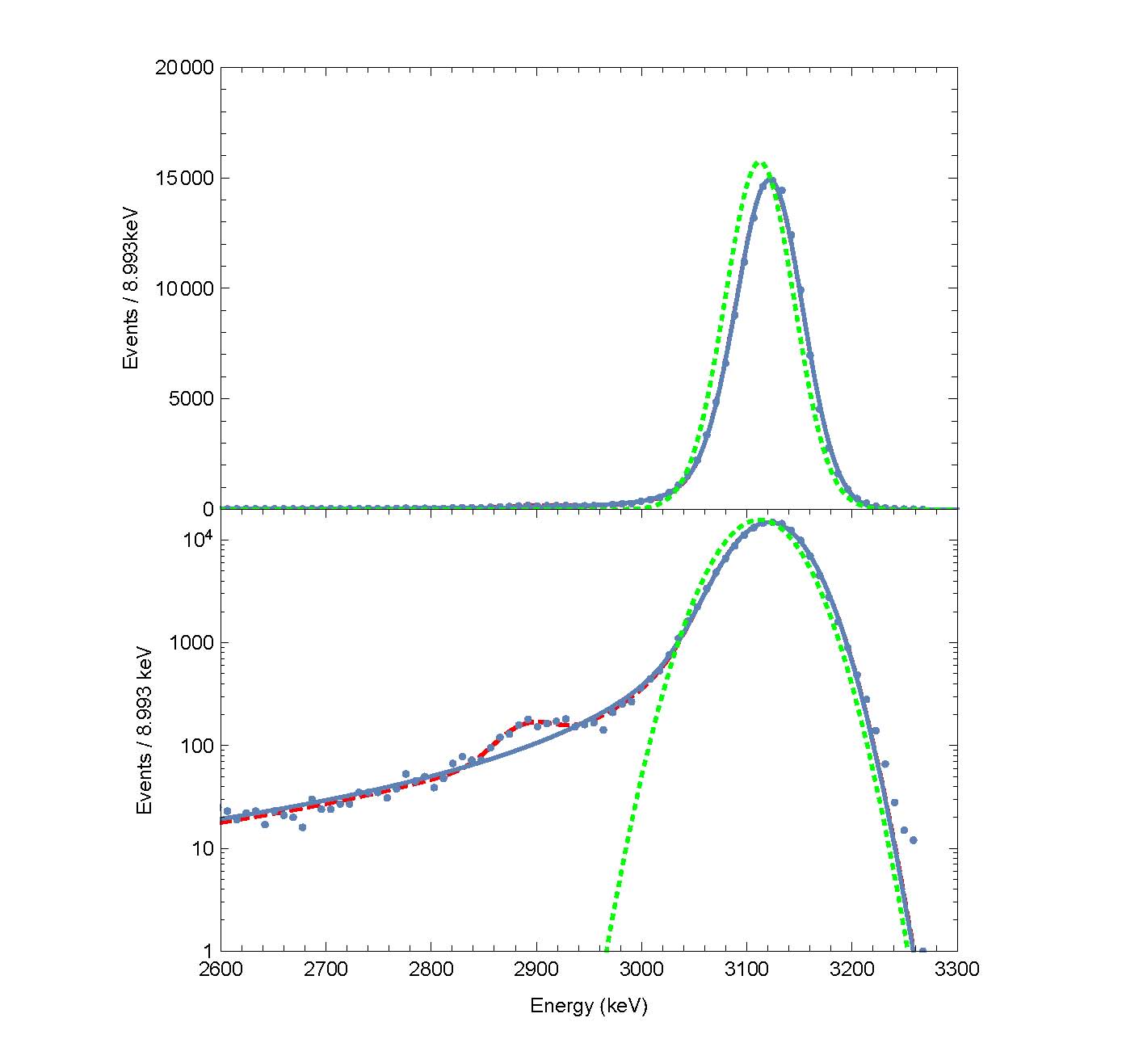}

\caption{Comparison of a $^{148}$Gd alpha energy spectrum with a $\sim 200\,\mu$g/cm$^2$ Au overcoat  (blue data points) compared to several models.  The SRIM MC output is represented here by the fit of Fig.\,\ref{fig:SRIMComareRWAuOveroat} with the green dotted line. The solid blue line is a five parameter fit to the data using Eqn.\,(\ref{eqn:OvercoatEnergyDistribution}). The red, dashed line is an eight parameter fit to the data, adding a Gaussian Distribution near 2900\,keV. The parameters for all the fits are listed in Table \ref{table:FitsToGoldOvercoat}.}
\label{fig:SRIMAuOvercoatDataGdSource}
\end{center}
\end{figure}

However, when both the SRIM MC data (as implemented here) or our intuitive prediction are confronted with data, neither predicts the shape of a $^{148}$Gd alpha energy spectrum after transmission through a
$\sim 200\,\mu$m thick gold overcoat, see Fig.\,\ref{fig:SRIMAuOvercoatDataGdSource}. Both SRIM and our prediction badly underestimate the tail of the energy distribution. Our model is flexible enough to reasonably fit the data, but the values of the parameters returned by the fit loose their meaning and do not correspond to their intuitive values. The model, and the SRIM MC data we generated, assume that the density of gold is the nominal value for solid gold metal (the ``stopping power"  in units of  (MeV\,cm$^2$/gm) appears in our model 
only multiplied by the density and therefor they are 100\% correlated). A possibility is that the gold overcoat of the source
is significantly less dense than the published Au density values \,\cite{HandbookOfChemistry}, throwing off the values in the fit from our intuitive starting point.
We did not run SRIM for lower Au densities  while maintaing a total amount of Au at $\sim200\,\mu$g/cm$^2$.

An interning feature in the spectrum is an apparent excess of events near $2900$\,keV. We are unable to explain this feature. We have characterized this excess by adding a  Gaussian distribution ($G(E)=G_0\exp[-[E-E^{\prime}_0]^2/2s^2]$) to our  intuitive prediction, and find a mean energy of $E^{\prime}_0=2892.5\pm2.7$\,keV. One possibility is that this peak comes from an unknown contamination in the $^{148}$Gd source. However, this energy does not correspond to any known isotope\,\cite{TableOfIsotopes} either as a contamination on the surface or with the energy degraded by $0\lesssim\delta E\lesssim70$\,keV from the Au overcoat. The alpha decay energy of $^{154}$Dy is 2872\,keV, or about 4.6 sigma away from the mean of the fitted peak (see Table \ref{table:FitsToGoldOvercoat}) were it a contamination on the surface of the gold overcoat. Its energy would be degraded  even lower by the Au overcoat were it commingled with the Gd source. An intriguing possibility is $^{149}$Gd, with an alpha energy of 3016\,keV, but we would expect the alpha from $^{149}$Gd to appear at 2940\,keV, also inconsistent with the data. Also, the half life of $^{149}$Gd is 9.28 days \,\cite{TableOfIsotopes}, and the amplitude of the peak is stable over months. Two other possibilities are a non-uniform distribution of Au over the source, or an inelastic reaction of the form Au(He,He$^{\prime}$)Au* but there is no corresponding structure in the $^{247}$Am peak (not shown). However, our uncoated, electroplated Gd-Am souce shows a similar structure at the equivalent, but slightly higher, energy. 

\begin{table}[htdp]
\caption{Naive and Fit parameters for Figs.\,\ref{fig:SRIMComareRWAuOveroat} and  \ref{fig:SRIMAuOvercoatDataGdSource}. The values in the first line are those used to create the SRIM MC data
set and the naive parameter inputs to Eqn.\,(\ref{eqn:OvercoatEnergyDistribution}). The second set of values is the fit of Eqn.\,(\ref{eqn:OvercoatEnergyDistribution}) to the MC data. Note the errors on the individual parameters are large, indicating too many free parameters for this approximately Gaussian distribution. The third set of values are the fit parameters of Eqn.\,(\ref{eqn:OvercoatEnergyDistribution}) to the $^{147}$Gd source data with a $\sim 200\,\mu$m/cm$^2$ Au overcoat shown in Fig.\,\ref{fig:SRIMAuOvercoatDataGdSource}. The fractional error values are much smaller than the previous case, but some of the parameter values have lost a simple interpretation with respect to the naive values of the first line of this table. The last two lines are again a  fit to Fig.\,\ref{fig:SRIMAuOvercoatDataGdSource}, but with the addition of a Gaussian distribution to account for the structure near $\sim 2900$\,keV. }
\begin{center}

\begin{tabular}{ccccccc}

  $A_{oc}$ &  $T_{\alpha}$  & $\sigma$ & $dE/dx$  & $R_{max}$&$\chi^2 $\\
  $(\times10^3)$ & (keV) &(keV) & (keV/$\mu$m) & ($\mu$m )& $\overline{ \text{DOF}}$ \\
\hline\hline
935 & 3182.8 &30.0 & 552.0  & 0.169 & - \\
\hline
$481\pm 282 $ &  $3138\pm 43$ & $31.1\pm1.4$   &  $144  \pm334$&$ 0.37\pm 0.58 $&0.49\\
\hline\hline
 & & & & & \\
$1282.0\pm3.4$ & $3132.8\pm 0.2$ &  $30.50\pm0.09$& $42.3\pm0.7$ & $19.3\pm 0.4$ & 4.4 \\
\hline\hline
 & & & & & \\
$1278\pm3.5$ & $ 3132.0\pm 0.2$ & $30.69\pm0.09$ & $39.1\pm 0.8 $ & $ 20.0\pm0.5 $& 3.0\\
\hline
 $G_0$ & $E^{\prime}_0$ & $s_0$ &  &  & \\
 $(\times10^3)$ & (keV) & (keV) & & & \\ 
 \hline
 $4.80 \pm 0.49 $ & $ 2892.5\pm2.7$ & $25.6\pm2.6 $  & & &\\
 \hline\hline
 \end{tabular}
\end{center}
\label{table:FitsToGoldOvercoat}
\end{table}%

\vspace{0.2in}
The goal of this paper was to provide a way to estimate radioactive contamination from degraded alpha spectra measured in neutrinoless double beta decay experiments, dark mater experiments or similar situations in background regions at lower energy  ($\lesssim$10\,MeV). We have shown it is possible to compute the the surface decay rate of a material uniformly contaminated with an alpha emitter to a few percent. We first verified could reproduce the intensity of a mixed, calibrated $^{148}$Gd \textendash$^{247}$Am to 99.8\% $\pm 0.7\%$\,(stat) $\pm1.6\%$\,(sys). The surface rate of the decay of $^{147}$Sm in a sample of Sm metal foil and the result agrees with our measurement at the level of  $102.3\%\pm 1.6\%\text{ (stat) }\pm 2.1\% \text{ sys}$. 
Alternately, if one has a material believed to be uniformly contemned with an alpha emitter, the model can be inverted and a measurement of the surface rate can be used to calculate the decay rate in the bulk. This technique is useful, for example, to quantify the bulk contamination is in the active part of a detector.

We have extended our intuitive model to the cases of an alpha emitter with an exponential distribution near the surface of a material and the case of a fixed overcoat above an alpha source.
We have shown that even with the parameter of our model set to intuitive values, we obtain excellent agreement between the sophisticated SRIM Monte Carlo calculations and our naive predications. When fit to the data, the $\chi^2$/DOF returned is typically $\lesssim 1$.  The model we describe is intuitive with parameters that are easily understood in an experimental context.
The formalism we introduced for emitters with an exponential distribution can easily be extended to any analytic distribution of either single or multiple alpha emitters.

However, neither our SRIM MC data (as generated here), nor our model can explain the measured energy spectrum of a mixed  $^{148}$Gd \textendash$^{247}$Am source with Au overcoat reported here, assuming the nominal density of gold.  Experiments to
improve the scientific case might include carefully controlled overlays (including both density and thickness) of other materials over a well calibrated source, or the controlled imbedding of alpha emitters versus distance into a base material and measuring the resulting surface rates of the alpha particles.

This paper is dedicated to the memory of Stuart Jay Freedman and Carl Frederick Gerhardt, Jr. We thank Eric (Rick) Norman for first pointing out the flat alpha energy spectrum of Samarium;
Sam Majier for pointing out reference\,\cite{AnalyticalAcceptance}; Brian Fujikawa for reviewing an early draft of this paper  and both Victor Gehman and Yuan Mei for useful discussions and help with\,\cite{Mathematica}.

This work was supported by the Director, Office of Science, Office of High Energy Physics, of the U.S. Department of Energy under Contract No. DE-AC02- 05CH11231 and The Richard W. Kadel Living Trust.


\end{document}